# A semiclassical 1D model of ultrafast photoisomerization reactions

V.A. Benderskii,[1] E.V. Vetoshkin,[1] E.I. Kats,[2,3] and H.P. Trommsdorff[2,4]

[1]Institute of Problems of Chemical Physics, RAS, 142432 Moscow Region, Chernogolovka, Russia

[2]Institut Laue-Langevin, F-38042, Grenoble, France

[3]L. D. Landau Institute for Theoretical Physics, RAS, Moscow, Russia

[4]Laboratoire de Spectrométrie Physique, Université Joseph-Fourier de Grenoble, CNRS (UMR5588), B.P. 87, F-38402 St. Martin d'Hères Cedex, France

**Abstract**

A simple model for ultrafast and efficient photoisomerization reactions is presented. In this model, the relevant region of the potential results from double crossing of two harmonic 1D diabatic potentials with opposite curvature. The eigenvalue problem is solved within a semiclassical approximation using the method of connection matrices, and the propagation of wave packets, expanded over these semiclassical eigenfunctions, is calculated. Within this model, fast double nonadiabatic transitions through both crossing points circumvent slow tunneling through the potential barrier. The extension to real systems makes this model relevant for the design of bi-stable photochromic materials.

# 1. Introduction.

Photo-induced isomerization reactions, taking place on ultrashort time scales ($< 10^{-12}$ s) and with high quantum yield ($> 0.1$), are the basis of many important processes: the vision process relies on this reaction in retinal-like molecules [1], and the response of many photochromic materials also involves isomerization reactions in bistable molecules [2-8]. The proper understanding and modeling of these reactions has therefore been a theme of ongoing interest but agreement between experiment and theoretical modeling leaves much to desire. Models presented about twenty years ago [9, 10] assume a ground electronic state potential energy surface (PES) with two deep minima and a single well excited state PES. These PES result from the avoided crossing of two ground state PES, describing each of the two isomers. Efficient and ultrafast conversion occurs when the splitting of these PES at the avoided crossing is not too large. For a large gap, isomerization occurs only subsequent to vibrational relaxation and will not be ultrafast. The case of a lower energy excited state PES, intersecting twice the ground state barrier was discarded, since the resulting adiabatic upper PES has two minima, separated by a barrier, and slow tunneling through this barrier was thought to limit the reaction rate.

In this communication we analyze in more detail such a model of double deep crossing. We show that when double nonadiabatic transitions are taken into account as an alternative mechanism to tunneling both the high efficiency and the high reaction rate of photoisomerization reactions can be explained. The vinylidyne-acetylene rearrangement through a hydrogen bridged structure is an example of a reaction involving a PES with an excited state double well potential, which can be associated with the double crossing of diabatic potentials [11, 12].

A further motivation of this work is that similar problems have not been studied by quantum mechanical methods. Only quantum dynamics simulations have been used to study ultrafast phenomena [13-16]. However, these calculations are intricate and the results are often difficult to interpret in terms of eigenstates and transition amplitudes of conventional quantum mechanics so that physical insight and a qualitative picture is lost. In this communication we present a simple model which enables us to obtain all results within standard quantum-mechanics of semiclassical accuracy. A full account, presenting the technical aspects of this development, is published elsewhere [17].

For various photoisomerization reactions, ultrafast conversion in the vicinity of conical intersections is often discussed as reaction mechanism and is supported by the results of quantum dynamics simulation. However, even for a 2D PES the quantum solution of the time dependent problem has not yet been obtained and the validity of quantum dynamics simulations for multidimensional systems with quasi-continuous spectra, when the density of states involved in the system evolution becomes very high, is not firmly established. For these reasons, we return to a much simpler model [9, 10] for which a semiclassically rigorous solution can be derived and results can be presented in the transparent form of traditional quantum mechanics. The price to pay for this simplification, is that our simple model cannot claim direct comparisons with isomerization reactions in multidimensional PES, and the role of the resonances and conical intersections is therefore beyond the scope of this paper.

In the following section 2, we present the model PES and outline the method. The eigenstates of this PES and the dynamics of wave packets are discussed in section 3. Finally, in Section 4 we conclude.

## 2. The model PES and summary of the method

Figure 1 shows the relevant part of the PES, in which the isomerization reaction is assumed to take place. The ground state minima are sufficiently deep so that the density of states in the intersection region of a *n*-dimensional PES is extremely high (in the of order $(E/\hbar\bar{\omega})^n$ where $\bar{\omega}$ is the average vibration frequency and E is energy gap). As the spectrum of excited vibrational states is near continuous, the ground state PES in this region can be replaced by a decaying potential. As a result, the diabatic PES (dashed lines) reduces to two intersecting parabolas with opposite curvatures. For the sake of simplicity, the absolute value of the curvature is taken to be the same in the calculations performed below, but the generalization to different curvatures, as shown in the figure, is straightforward. In order to emphasize the difference between isomers, the two PES are displaced horizontally by $\pm X_0$ so that coupling of these PES produces an asymmetric adiabatic PES presented by the full lines. The adiabatic potentials are given in the following 2×2 matrix form:

$$V = \begin{pmatrix} -1+(X+X_0)^2 & u_{12} \\ u_{12} & 1-(X-X_0)^2 \end{pmatrix} \qquad (1)$$

The semiclassical parameter, γ, is defined as the squared ratio of a characteristic distance, *a*, given here by the half distance between the crossing points in the symmetric, case $X_0 = 0$, over the zero-point amplitude:

$$\gamma = \frac{m\Omega a^2}{\hbar} \qquad (2)$$

where $\Omega$ is the dimensional frequency of the potentials, energies are scaled by $\frac{1}{2}m\Omega^2 a^2 = \frac{1}{2}\gamma\hbar\Omega$ and distances by *a*. $\gamma$ is assumed to be sufficiently large ($\approx 10$) to assure a high accuracy of the semiclassical approximation. Within the model PES, Eq. (1), the eigenvalue problem is solved using the generalized semiclassical approach developed in the previous papers [18-20] and the evolution of two types of wave packets, as indicated in Fig. 1, is calculated.

The standard semiclassical approach to the eigenvalue problem is based on the formalism of connection matrices [21] for the turning points of first and second order and shift matrices between these points. In order to generalize this approach to nonadiabatic problems, connection matrices for crossing points must be incorporated into the set of connection matrices for turning points characterizing an arbitrary 1D potential. The simplest model of the crossing of linear potentials with coordinate independent coupling was introduced by Landau and Zener for predissociation reactions (see for example [22]). This model is the simplest topological element, which enables us to construct the connection matrix for a crossing point. This matrix has been derived recently for the whole energy region [18-20], ranging from regions below the top of lower adiabatic potential to above the minimum of the upper potential. The accuracy of the semiclassical approach regarding crossing point problems has been discussed earlier [19].

Once the set of connection and shift matrices characterizing a given potential is known, the total transformation matrix, which determines the asymptotic behavior of wave functions at $\pm\infty$, is written as scalar product of the above matrices and the quantization rules are derived as follows: wave functions of bound states must vanish at $\pm\infty$, while wave functions of quasi-stationary states do not contain ingoing waves from $\pm\infty$ to the interaction region. This procedure is applicable for the combination of any two 1D diabatic potentials.

**3. Eigenstates and wave packet dynamics.**

In the diabatic ($u_{12} \to 0$) and adiabatic ($u_{12} \to \infty$) limits, the eigen-spectra of bound states are found trivially, corresponding to those of a harmonic oscillator and an asymmetric double-well potential, respectively. For finite non-adiabatic couplings, $u_{12}$, we apply the techniques described above. The results, complex eigenvalues versus $u_{12}$, are presented in Figs. 2 and 3 (for $\gamma = 12$ and $X_0 = 0.1$). Simple inspection of Figs. 2 and 3 shows that there are two types of states in the upper adiabatic potential: purely decaying states (D states, represented by dashed lines in Fig. 2), and quasi-stationary states (Q states, full lines in Fig. 2), quasi-localized in the L ($Q_L$) or R ($Q_R$) wells. In the diabatic limit, the D states are stable, and the decay rate of these states grows monotonically with increasing adiabatic coupling (Fig. 3a). For sufficiently strong coupling ($u_{12} \geq 1 - 2$) the decay becomes barrierless and quasi-immediate. In the over-barrier region exist the delocalized Q-states. In contrast to D-states, localized and delocalized Q-states are stable in the energy window below the maximum of the adiabatic potential barrier, $V^{\#}$. These states are stable in the both limits ($u_{12} \to 0$ and $u_{12} \to \infty$), their decay reaching a maximum at intermediate values of $u_{12}$ (Fig. 3b). The decay rate of Q states becomes maximal for $n' = 30$, corresponding to energies of $E \approx 2V^{\#}$, decreasing sharply at higher energies.

The time evolution was calculated for two Gaussian wave packets with positions, I and II, as indicated in Fig. 1. For these positions, the energy of the left branch of upper adiabatic of the PES equals the energy of zero point level of the left well and $2V^{\#}$, respectively. Their spectral expansion can be represented as follows:

$$\Psi(X,t) = \sum_n a_n \exp\left(i\frac{E_n + i\Gamma_n}{\hbar}t\right)\Phi_n(X) \qquad (3)$$

$\Phi_n$ are the eigenfunctions of the adiabatic PES and $E_n + i\Gamma_n$ the corresponding complex eigenvalues. The expansion coefficients, $a_n$, are characterized by a Gaussian width of $\delta_0 = \gamma^{-\frac{1}{2}} = 0.289$ ($\gamma = 12$) for both wave packets. At least 30 eigenstates were included in expansion (3) in order to assure reasonable numerical accuracy.

The time evolution of wave packets I and II has been calculated in the wide range of coupling constants from the diabatic ($u_{12} = 0$) to the adiabatic ($u_{12} \gg 1$) limit. The evolutions are shown in Figs. 4 and 5. In the diabatic limit, wave packet I spreads over the wide lower diabatic well and recovers near the turning points with the residual amplitude (Fig; 4a). As the coupling increases, the wave packet amplitude decreases due to the decay of Q-states (Fig. 4b, c). In the adiabatic limit (Fig. 4d), packet I is localized in the left well and its decay rate becomes small, since nonadiabatic transitions are suppressed. The behavior of wave packet II is similar to that of packet I not only in the diabatic limit, but also in the wide range of intermediate coupling (Fig. 5a, b). In contrast to packet I, packet II is delocalized in the adiabatic limit due to fast overbarrier transitions. The similarity of the dynamics of packets I and II in the wide intermediate range of coupling is the result of double nonadiabatic transitions.

The appearance of almost all wave packet density at $t = 0.25$ in the classically forbidden region of the upper adiabatic potential, seems at first sight surprising. However, the barrier exists only in the adiabatic limit, while the system is strongly nonadiabatic in the intermediate region studied here. Due to this nonadiabaticity, an alternative reaction path of L↔R conversion appears.

This path involves a nonadiabatic transition in the L well through the left crossing point from the upper to the lower PES, displacement in this lower PES to the right crossing point and return to the upper PES through the right crossing point. This path involves two nonadiabatic transitions and avoids tunneling through the barrier between adiabatic L and R states.

The quantum yield of isomerization in n-th quasistationary state which exhibits two channel decay in L- and R-wells is defined by the ratio of probability flows through the left and right crossing points $X_L, X_R$:

$$Y_{LR}^{(n)} = \frac{\Gamma_R^{(n)}}{\Gamma_L^{(n)} + \Gamma_R^{(n)}} \quad , \quad \Gamma_{L(R)}^{(n)} = \Psi_n^* \frac{d\Psi_n}{dX} - \Psi_n \frac{d\Psi_n^*}{dX} \bigg|_{X=X_{L(R)}} \tag{4}$$

For a symmetric potential, $X_0 = 0$, this quantum yield always equals ½. In the asymmetric potential, the quantum yield decreases as a function of $u_{12}$ and reaches zero in the adiabatic limit due to the localization of adiabatic wave functions in the L or R wells. With increasing asymmetry, the region where $Y_{LR}$ drops is shifted towards smaller values of $u_{12}$. However, in the intermediate region ($u_{12}$ = 1-3), high values of quantum yield ($Y_{LR} \geq 0.1$) are preserved even when the asymmetry becomes comparable to the characteristic frequency. Moreover, for states above the barrier with $E > 1.5\, V^{\#}$, $Y_{LR} \approx 0.5$. Our model, therefore, provides a basis for both experimental findings, namely the high quantum yield combined with ultrafast dynamics of photoisomerization reactions.

## 4. Conclusions

To our knowledge, a study of double non-adiabatic transitions, as discussed here, has not been carried out previously. We could show here that the method of connection matrices works well for this problem, even in the range of intermediate coupling strength, where real and imaginary parts of the complex eigenvalues are of comparable magnitude. This is an important result, since standard known methods to solve such a problem, fail. A direct numerical diagonalization of the Hamiltonian (although feasible) becomes prohibitive in terms of computer time.

The specific model discussed in this paper allows to follow the wave packet evolution in terms of complex energy eigenvalues and transition amplitudes. Indeed, the expansion, Eq. (3), in the intermediate coupling region contains oscillator functions with large quantum numbers and only an exponentially small contribution from the continuous spectrum of wave functions. As the decay rates for states with different quantum numbers differ, the dynamics do not follow a simple exponential decay law. In contrast, in the adiabatic limit, oscillator functions with small quantum numbers are mainly represented in the initial wave packet. In principle, such theoretical predictions, regarding the fast time evolution of the initial state, can be tested by femtosecond pump-probe experiments [23, 24].

In conclusion, we note briefly the way in which our model can be extended to multidimensional PES as determined in quantum chemical calculations. In multidimensional PES, the 1D path must be replaced by the minimum action path (MAP), which obeys classical equations of motion. If resonances and bifurcations of the MAP are absent, the total spectrum contains ladders of longitudinal quasistationary and transverse slightly anharmonic states, characterized by well defined quantum numbers. Within the adiabatic approximation, the dynamics of longitudinal states is described in terms of an effective 1D potential. The dynamics of transverse modes follows adiabatically the evolution of longitudinal states in the time scale determined by transition matrix elements, which are proportional to coupling constants between transverse and longitudinal coordinates. In the resonance regions, longitudinal and transverse motions are strongly mixed, so that the initial evolution involves a fast redistribution between the pure states. Although this redistribution is beyond the present 1D model, the subsequent dynamics of mixed states can be described by introducing an effective 1D potential, obtained by a coordinate transformation from the basis of initial pure states to a new basis, in which resonance terms are eliminated. As the pure longitudinal states, the mixed states are quasistationary. Their dynamics is similar to the one described by the model discussed here and involves ultrafast relaxation and L↔R transitions. The mixing of longitudinal and transverse states of antisymmetric vibrations, taking place in the vicinity of a conical intersection at energies above the top of the cone top, is considered to be the mechanism of ultrafast isomerization reactions [25]. In order to take this mixing into account within an extension of our model, it will be necessary to construct the connection matrices for the conical intersections. The cis-trans isomerization of ethylene-like molecules is associated with a hindered rotation about C=C bond where the minima in the exited electronic state are related to the twisted geometry of a double bond. This system, studied earlier within the model of conical intersections [13-16,26,], is an interesting object for a comparison of double crossing and conical intersection models.

**Acknowledgements**

This work would have not been possible without stimulating discussions with T. Ziman. One of us (E.K.) is thankful to INTAS Grant (under No. 01-0105) for partial support, and V.B. and E.V are indebted to CRDF Grant RU-C1-2575-MO-04.

References

1. F. Gai, K. C. Hasson, J. C. McDonald, P. A. Anfinrud, Science **279** (1998) 1886
2. H. Dürr, H. Bouas-Laurent, *Photochromism: Molecules and Systems*, Elsevier, Amsterdam 2003.
3. L. Song, M. A. El-Sayed, J. K. Lanyi, Science **261** (1993) 891.
4. S. Logunov, M. A. El-Sayed, l. Song, J. Phys. Chem. **100** (1996) 2875.
5. K. Lednev, T-O. Ye, R. E. Hester, J. N. Moor, J. Phys. Chem. **100** (1996) 13333.
6. S. Lochbrunner, W. Fuss, W. E. Schmid, L-E. Kompa, J. Phys. Chem. **102** (1998) 9334.
7. W. Fuss, W. E. Schmid, S. A. Trushin. J. Chem. Phys. **112** (2000) 8347.
8. J. Ern, A. T. Bens, H.-D. Martin, K. Kuldova, H. P. Trommsdorff, C. Kryschi. J. Phys. Chem. **A106** (2002) 1654.
9. L. Salem, Science **191** (1976) 822.
10. J. Michl, V. Bonacic-Koutecky, Electronic aspects of organic chemistry, Wiley, New York, 1990.
11. T. Carrington, Jr., L. M. Hubbard, H. F. Schaefer III, W. H. Miller, J. Chem. Phys. **80** (1984) 4347.
12. M. M. Gallo. J. Am. Chem. Soc. **112** (1990) 8714.
13. H. Köppel, Chem. Phys. **77** (1983) 359.
14. U. Manthe, H. Köppel, J. Chem. Phys. **93** (1990) 1658.
15. M. Ben-Nun, T. J. Martinez, Chem. Phys. Lett. **298** (1998) 57.
16. M. Ben-Nun, F. Molhar, H. Lu, L. C. Phillips, T. J. Martinez, K. Shulten, Farad. Disc. **110** (1998) 447.
17. V. A. Benderskii, E. V. Vetoshkin, E. I. Kats, to be published
18. V. A. Benderskii, E. V. Vetoshkin, E. I. Kats, JETP **97** (2003) 232.
19. V. A. Benderskii, E. V. Vetoshkin, E. I. Kats, Phys. Rev. **A 69** (2004) 062508.
20. V. A. Benderskii, E. V. Vetoshkin, E. I. Kats, JETP Lett. **80** (2004) 493.
21. J. Heading, An Introduction to Phase-Integral Methods, Wiley - Interscience, London, 1962.
22. L. D. Landau, E. M. Lifshitz, Quantum Mechanics, Pegamon Press, New York, 1965.
23. G. R. Fleming, Chemical applications of ultrafast spectroscopy, Oxford University Press, London, 1986.
24. S. Mukamel, Principles of nonlinear optical spectroscopy, Oxford University Press, London, 1995.
25. H. Köppel, W. Domcke, L. S. Cederbaum. Adv. Chem. Phys. **57** (1984) 60.
26. M. Ben-Nun, T. J. Martinez, Chem. Phys. **259** (2000) 237.

27. W. Domcke, D. R. Yarkony, H. Köppel (Eds.), Conical Intersections: Electronic Structure, Dynamics and Spectroscopy (World Scientific, Singapore, 2004).

Figure captures.

Figure 1.

Diabatic (dashed line) and adiabatic (solid line) potentials of the ground and excited states in the double crossing region (the minima of the ground electronic state are not indicated, see text). The horizontal arrows indicate L to R isomerization. Positions I and II, at which wave packets are created, are indicated (see text).

Figure 2.

Real parts of the eigenvalues as functions of adiabatic coupling strength, $u_{12}$, in a potential with $X_0 = 0.1$, and $\gamma = 12$. Quantum numbers, $n'$, of the harmonic diabatic potential are indicated (for clarity only even numbers) at the left, while on the right side those of $Q_R$- and $Q_L$-states of the upper adiabatic potential are shown. D-states are marked by dashed lines up to values of $u_{12}$, for which the imaginary part of eigenvalues exceeds the energy spacings. The evolution of the stationary points, $V_+^{L(min)}, V_+^{R(min)}, V_+^{(max)}$, of the upper adiabatic potential are shown by dash-dotted lines.

Figure 3.

Decay rates of D- and Q-states in a potential with $X_0 = 0.1$, and $\gamma = 12$ as a function of $u_{12}$ (panels a and b, respectively). The quantum numbers relate to the diabatic states as indicated in Fig 2. Some intermediate levels with the similar behavior are not indicated.

Fig.4. Evolution of wave packets, created in position I in potentials characterized by $X_0 = 0.1$ and $u_{12} = 0.1$ (a), 1.2 (b), 2.5 (c), 5.0. The upper adiabatic potential is shown by a dashed line. Time, measured in the scale of the oscillation period in the diabatic potential, increases from top to bottom as $t = 0, 0.25, 0.5, 1$. For times $t > 0$, the amplitude of the wavepacket is multiplied by 2 in panel (b) and 4 in panel (c).

Fig. 5. Dynamics of the wave packet created in position II in potentials characterized by $X_0 = 0.1$ and $u_{12} = 0.1$ (a), 1.7 (b), 3.0 (c). The upper adiabatic potential is shown by a dashed line. Time, measured in the scale of the oscillation period in the diabatic potential, increases from top to bottom as $t = 0, 0.25, 0.5, 1$. For times $t > 0$, the amplitude of the wavepacket is multiplied by 4 in panel (b).

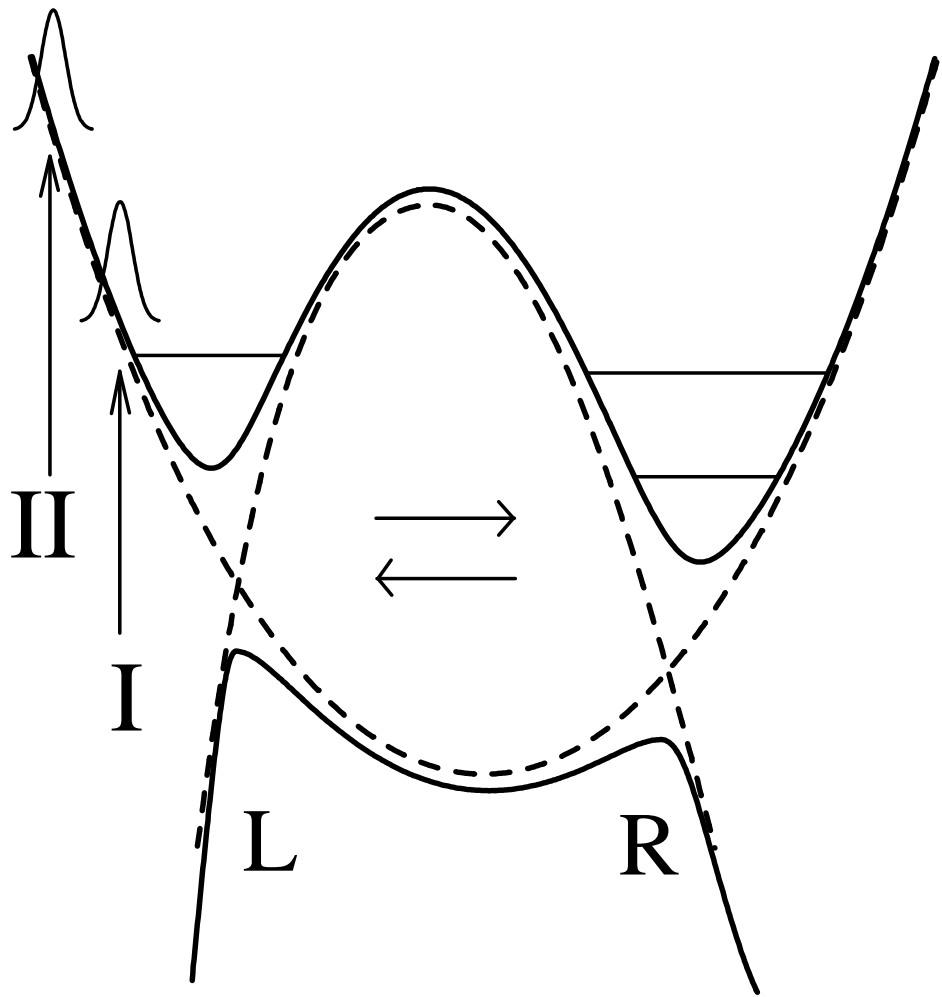

Fig. 1

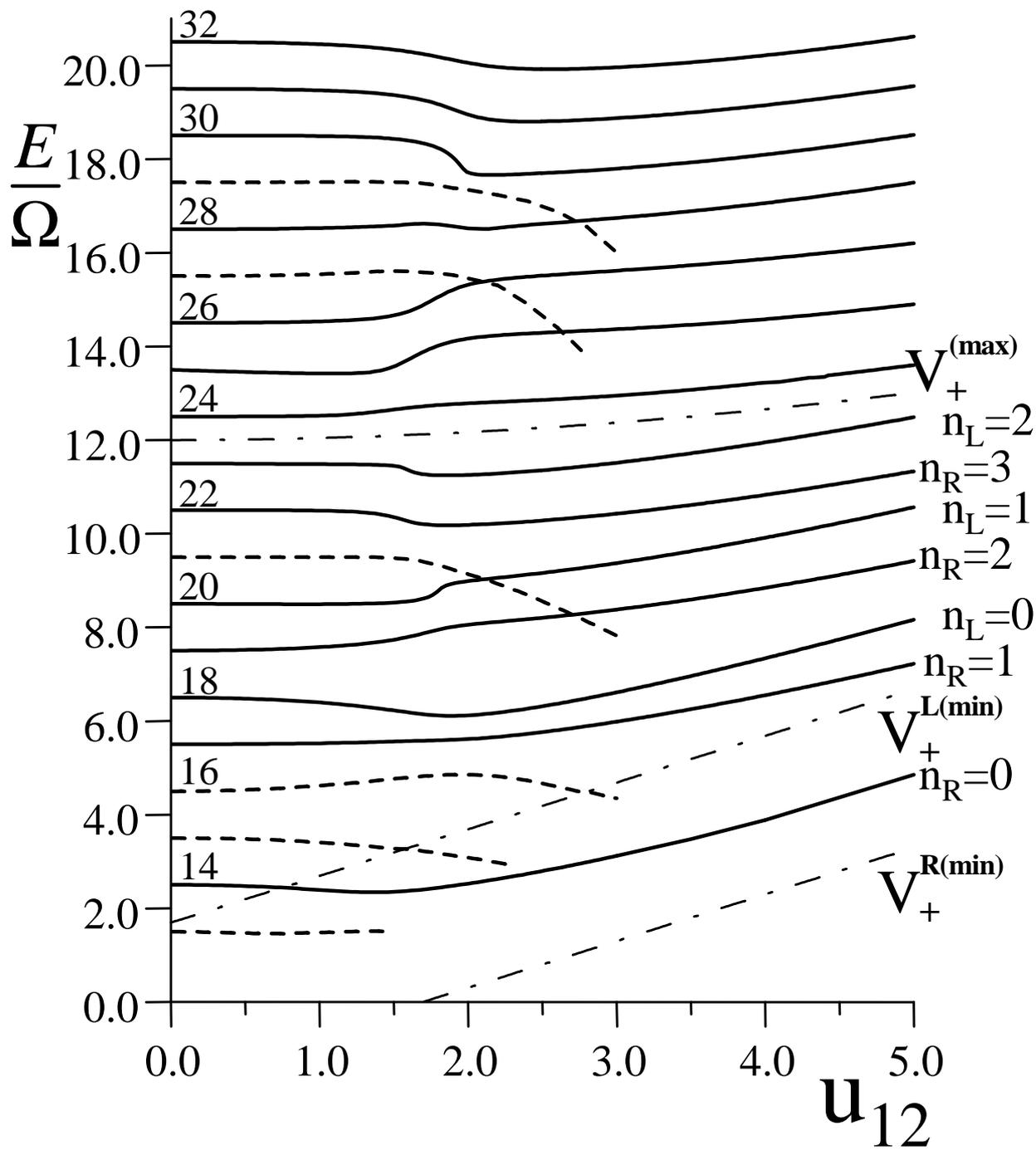

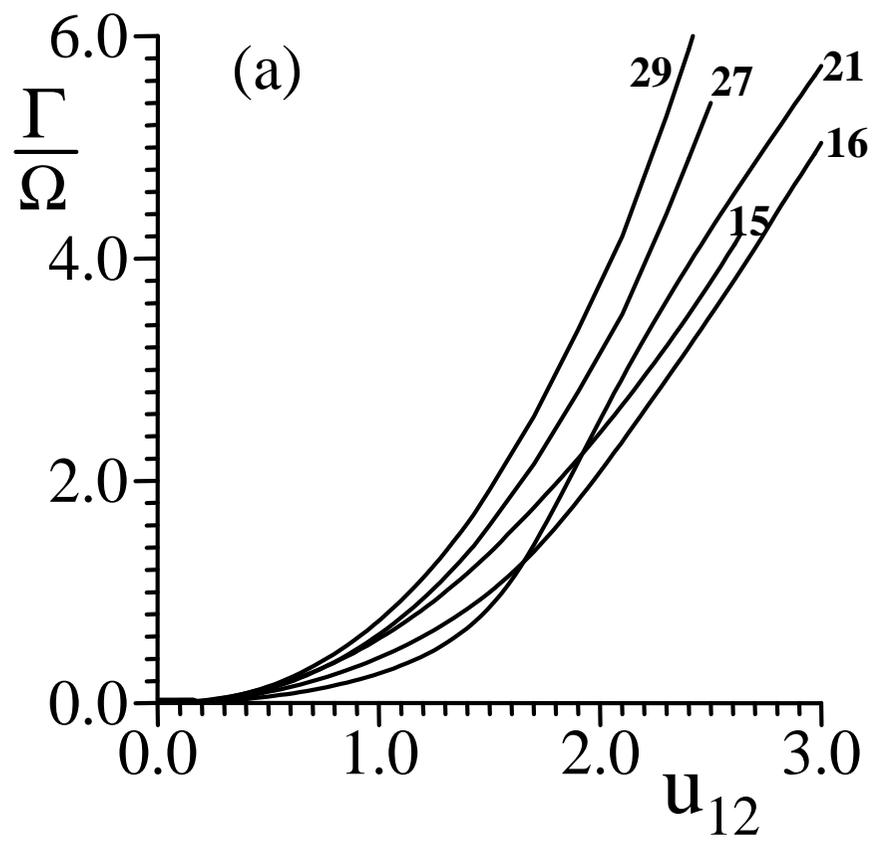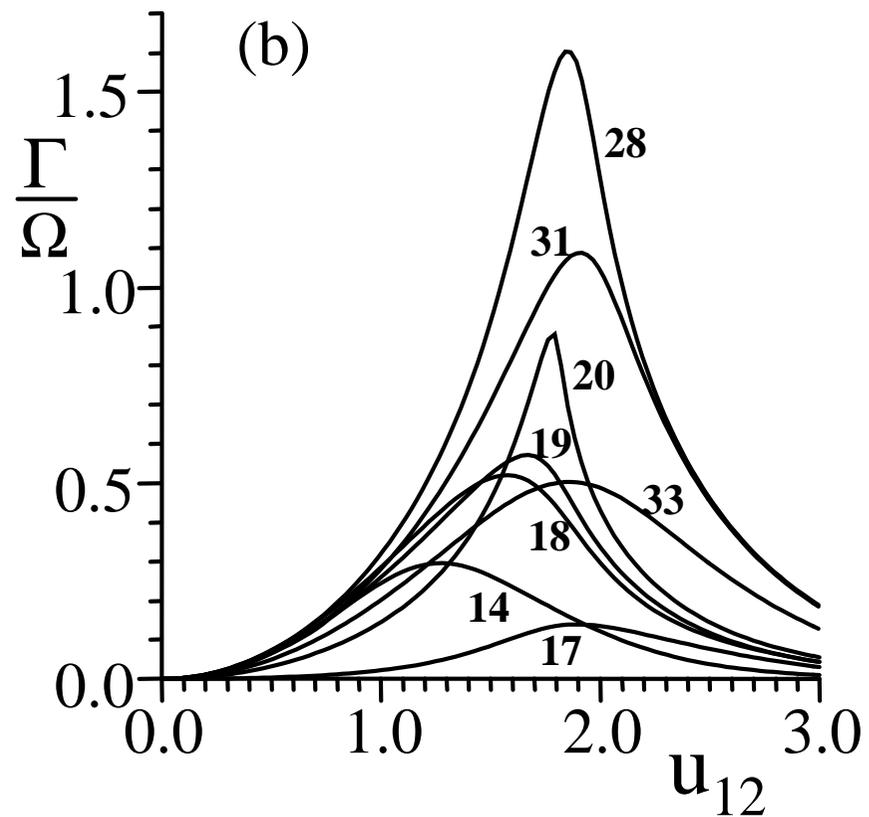

Fig. 3

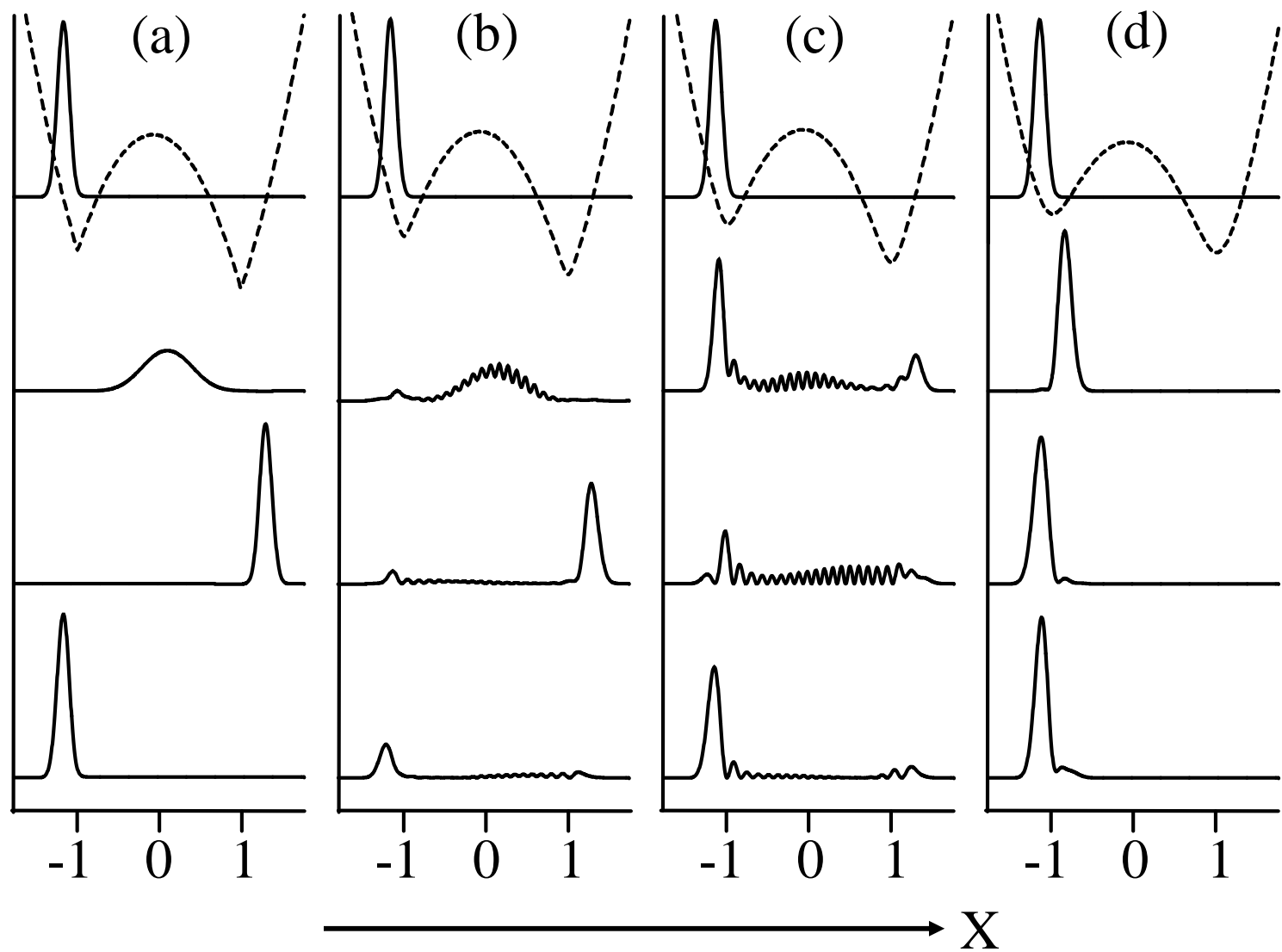

Fig. 4

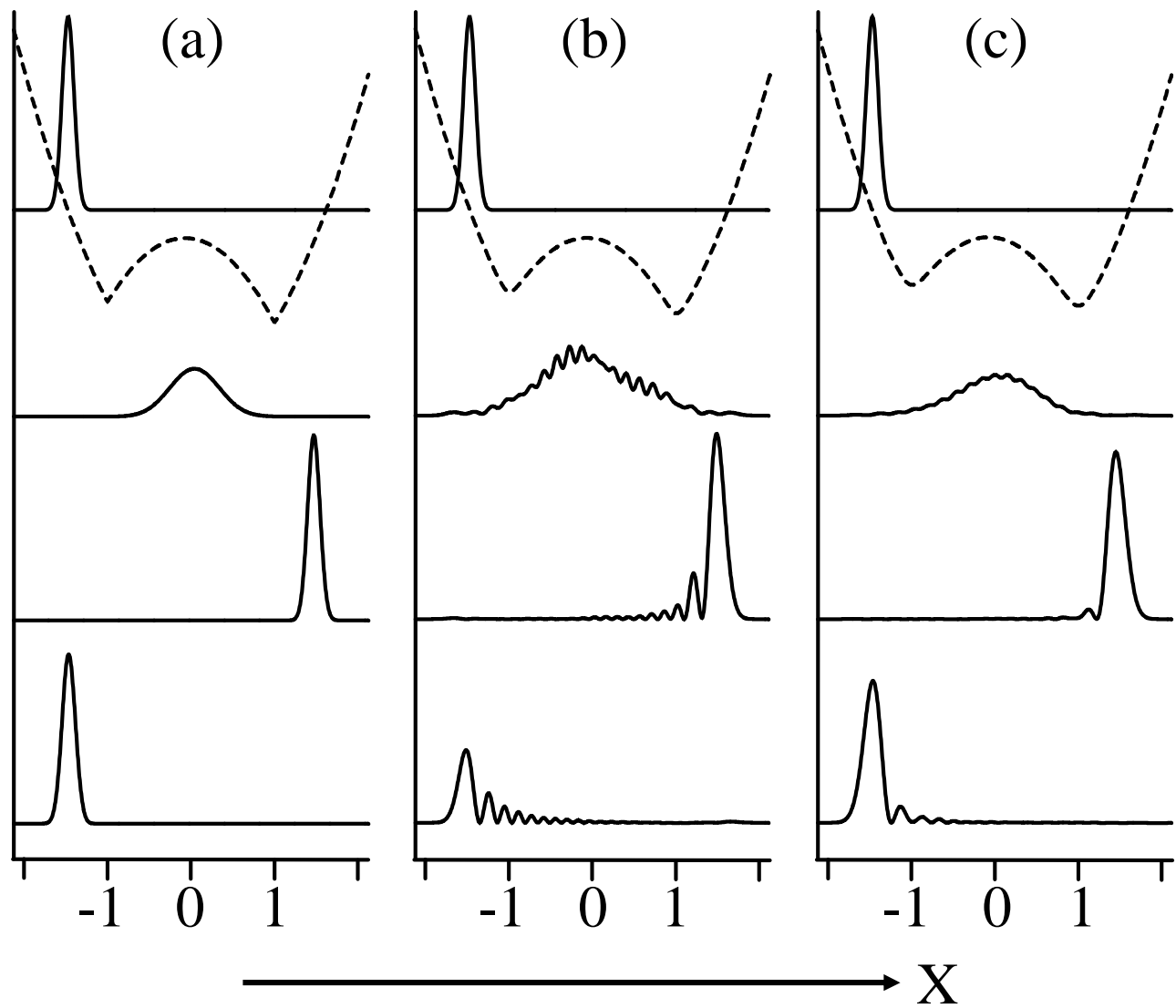

Fig. 5